\def\bar{\overline}

\def\*{\star}
\def\[{\left[}
\def\]{\right]}
\def\({\left(}
\def\){\right)}

\def\zb{{\bar{z} }}
\def\frac#1#2{{#1 \over #2}}
\def\inv#1{{1 \over #1}}
\def\half{{1 \over 2}}
\def\d{\partial}

\def\vev#1{\langle #1 \rangle}

\def\2pi{\hbox{$2\pi i$}}

\def\dsl{\raise.15ex\hbox{/}\kern-.57em\partial}
\def\Dsl{\,\raise.15ex\hbox{/}\mkern-.13.5mu D}

\def\ga{\gamma}		\def\Ga{\Gamma}

\def\al{\alpha}
\def\ep{\epsilon}
	
\def\de{\delta}		\def\De{\Delta}
		
\def\sig{\sigma}	

\def\CJ{{\cal J}}		
		\def\CO{{\cal O}}

\font\numbers=cmss12
\font\upright=cmu10 scaled\magstep1
\def\stroke{\vrule height8pt width0.4pt depth-0.1pt}
\def\topfleck{\vrule height8pt width0.5pt depth-5.9pt}
\def\botfleck{\vrule height2pt width0.5pt depth0.1pt}
\def\Zmath{\vcenter{\hbox{\numbers\rlap{\rlap{Z}\kern
		0.8pt\topfleck}\kern
		2.2pt \rlap Z\kern 6pt\botfleck\kern 1pt}}}
\def\Qmath{\vcenter{\hbox{\upright\rlap{\rlap{Q}\kern
                   3.8pt\stroke}\phantom{Q}}}}
\def\Nmath{\vcenter{\hbox{\upright\rlap{I}\kern 1.7pt N}}}
\def\Cmath{\vcenter{\hbox{\upright\rlap{\rlap{C}\kern
                   3.8pt\stroke}\phantom{C}}}}
\def\Rmath{\vcenter{\hbox{\upright\rlap{I}\kern 1.7pt R}}}
\def\Z{\ifmmode\Zmath\else$\Zmath$\fi}
\def\Q{\ifmmode\Qmath\else$\Qmath$\fi}
\def\N{\ifmmode\Nmath\else$\Nmath$\fi}
\def\C{\ifmmode\Cmath\else$\Cmath$\fi}
\def\R{\ifmmode\Rmath\else$\Rmath$\fi}
\def\cadremath#1{\vbox{\hrule\hbox{\vrule\kern8pt\vbox{\kern8pt
			\hbox{$\displaystyle #1$}\kern8pt}
			\kern8pt\vrule}\hrule}}

\def\presentation{
\voffset -.50in   %\voffset -1.05in
\hoffset -.19in
\oddsidemargin 0in \evensidemargin 0in
\marginparwidth .75in \marginparsep 7pt \topmargin 0in
\headheight 12pt \headsep .25in
\footheight 18pt \footskip .35in
\textheight 9.5in \textwidth 6.5in
\columnsep 10pt \columnseprule 0pt }
\def\debut{ \begin{eqnarray} }
\def\fin{ \end{eqnarray} }
\def\non{ \nonumber }
\documentstyle[12pt]{article}
\presentation
\begin{document}
\rightline{SPhT-95-006}
\rightline{hep-th/9501087}
\vskip 1cm
\centerline{\LARGE On The Random Vector Potential Model}
\bigskip
\centerline{\LARGE In Two Dimensions.}
\vskip 1cm
%
%\centerline{\large Olivier Babelon }
%\centerline{Laboratoire de Physique Th\'eorique et Hautes
%Energies \footnote[1]{\it Laboratoire associ\'e au CNRS.}}
%\centerline{ Universit\'e Pierre et Marie Curie, Tour 16 1$^{er}$
%\'etage, 4 place Jussieu}
%\centerline{75252 Paris cedex 05-France}
%
\vskip1cm
\centerline{\large  Denis Bernard
\footnote[2]{Member of the CNRS} }
\centerline{Service de Physique Th\'eorique de Saclay
\footnote[3]{\it Laboratoire de la Direction des Sciences de la
Mati\`ere du Commissariat \`a l'Energie Atomique.}}
\centerline{F-91191, Gif-sur-Yvette, France.}
\vskip2cm
Abstract.\\
The random vector potential model describes massless fermions
coupled to a quenched random gauge field. We study its abelian
and non-abelian versions. The abelian version can be completely
solved using bosonization. We analyse the non-abelian model using
its supersymmetric formulation and show, by a perturbative
renormalisation group computation, that it is asymptotically
free at large distances. We also show that all the quenched
chiral current correlation functions can be computed exactly,
without using the replica trick or the supersymmetric formulation,
but using an exact expression for the effective action for any sample
of the random gauge field. These chiral correlation functions
are purely algebraic.
\vfill
\newpage
%
%%%% DEBUT  %%%%%%%%%%%%%%%%%%%%%%%%%
%
%
Two dimensional random systems are much less understood than
their pure companions. It is not clearly
understood how (if possible) conformal field theory
techniques can be applied to random systems.
For example, what kind of (probably non-unitary)
conformal field theories
could describe the infrared fixed points of two dimensional
random systems is still an open question.

Besides the familiar study of the effects of
disorder on phase transitions \cite{DoDo,CaLu,Lu,Sh}, there has been a
renewed interest in two dimensional
random systems in connection with the quantum Hall
phase transition \cite{Exp,Pr,Zi}, or in connection
with models of randomly pinned
flux lines in superconductors \cite{Fi}.

Obviously, we will not give a complete answer to the question
formulated above. However, the models we will study are simple
enough, and possess enough symmetries,
to allow for exact computations which do not rely on
the replica trick or on the supersymmetric formulation.
Using this direct computation, we derive some informations
on the operator product algebra governing the infrared
behavior of these random models. Although our result are
only partial results, we hope they provide an information
useful enough to start answering the question.

The abelian version of the models
was recently introduced for analysing the quantum Hall
transition in ref.\cite{ShLu}.
The non-abelian version is similar to the model introduced in
\cite{Tet} for describing the effects of impurities interacting
with fermions close to the Fermi surface in $2+1$ dimensional systems.
However, as explained in the Appendix, it seems that our conclusions
are in disagreement with some of those of ref.\cite{Tet}.

\def\Asl{\raise.15ex\hbox{/}\kern-.67em A}
\def\zb{ {\bar z} }

\section{The model.}

\bigskip
\noindent {\it i) Definition of the model.}

The model describes $N$ massless Dirac fermions minimally coupled to a random
non-abelian gauge field with euclidean action~:
$S=\inv{\pi}\int d^2x\bar \Psi(i\dsl+\Asl)\Psi$. It is useful to
introduce the complex coordinates $z=x+iy$, $\bar z = x-iy$
and the components of the fermions $(\bar \psi^j_+,\psi^j_+)$ and
$(\bar \psi_{-;j},\psi_{-;j})$, $j=1,\cdots,N$. The action then becomes~:
\debut
S[A]=\int \frac{d^2z}{\pi}
\({\psi_{-;j}\( \d_\zb \de^j_k + A^j_{\zb,k}\)\psi^k_+ +
\bar \psi_{-;j}\( \d_z \de^j_k + A^j_{z,k}\)\bar \psi^k_+ }\)
\label{action}
\fin
where $A^j_{z,k}=i\sum_a A^a_z (t^a)^j_k$, with $(A^a_z)^*=A_\zb^a$,
is the gauge field.
Here the hermitian matrices $t^a$ form the $N$-dimensional representation
of $U(N)$. We denote by $f^{abc}$ the U(N) structure constants~:
$[t^a,t^b]=if^{abc}t^c$. The Dirac fermions take values in this
$N$-dimensional representation.

The gauge field is assumed to be a quenched variable with a Gaussian
measure~:
\debut
P[A] = \exp\[{ - \inv{\sig}\int~\frac{d^2z}{\pi}~\sum_a A^a_z A^a_\zb }\]
\label{pa}
\fin

We will be interested in computing the quenched average of the
correlation functions of the currents $J^a_\mu= (\bar \Psi
\gamma_\mu t^a \Psi)$. Explicitely, their components are~:
\debut
J^a_z &=& \psi_{-;j} (t^a)^j_k \psi^k_+ \label{current}\\
J^a_\zb &=& \bar \psi_{-;j}  (t^a)^j_k \bar \psi^k_+ \non
\fin
A convenient way of doing it consists in introducing sources
for the currents.  This amounts to shift the random field $A$
by an external sources $a_{\rm source}$~:
$A \to \cal A= A + a_{\rm source}$. We are then interested
in computing the average of the logarithm of the
partition functions with sources~:
\debut
 \Ga[a_{\rm source}]= {\bar {W[\cal A]}} = {\bar {\log Z[\cal A]}}
\label{effec}
\fin
with $\cal A= A + a_{\rm source}$ and
\debut
Z[A]= e^{W[A]} = \int D\Psi \exp\({ - S[A] }\)=
 Det[i\dsl+\Asl] .\label{part}
\fin
$W[A]$ is the generating function for the connected Green function~:
\debut
\vev{J^a_\mu(x)J^b_\nu(y)\cdots}^c_A = \pi^{2+\cdots}
\frac{\de W[A]}{\de A^b_\nu(y) \de A^a_\mu(x) \cdots } \non
\fin

\bigskip
\noindent {\it ii) The abelian case.}

Before plunging into the non-abelian case,
let us first present a simple
way of solving  the abelian case which corresponds to $N=1$.
This model was introduced
in ref.\cite{ShLu} in connection with the quantum Hall
transition, and studied there using the replica trick.
For $N=1$, the action (\ref{action}) can be bosonized.
Using the standard rules, $\bar \Psi\ga_\mu\Psi = \ep_{\mu\nu}
\d_\nu\Phi$, and $\bar \Psi i\dsl \Psi = \half (\d_\mu\Phi)^2$,
one finds the bosonic form of the action  (\ref{action})~:
\debut
S^{N=1}[\Phi,A]= \int\frac{d^2x}{\pi}\({
\half (\d_\mu\Phi)^2 + iA_\mu \ep_{\mu\nu}\d_\nu\Phi }\).\non
\fin

Since it is a gaussian model, it can be easily solved
without using the replica trick.
Let us introduce the Hodge decomposition of $A_\mu$~:
$A_\mu=\d_\mu\xi + \ep_{\mu\nu}\d_\nu\eta$.
The fields $\xi$ and $\eta$ decouple in the measure (\ref{pa})~:
\begin{eqnarray}
P[\xi;\eta]=
 \exp\[{ - \inv{\sig}\int~\frac{d^2x}{\pi}~
 \({(\d_\mu\xi)^2 + (\d_\mu\eta)^2}\) }\] \label{meseta}
\end{eqnarray}

The action $S^{N=1}[\Phi,A]$ is independent of $\xi$ and
therefore the field $\xi$ is irrelevant. This fact was expected
since the field $\xi$ represents a pure gauge whereas the physically
relevent quantity is the field strengh $F=\ep_{\mu\nu}\d_\mu A_\nu
= (\d_\mu\d_\mu) \eta$. Moreover, the field $\eta$ can be
absorbed into a translation of $\Phi$~:
\debut
S^{N=1}[\Phi,A_\mu=\ep_{\mu\nu}\d_\nu\eta]=
S^{N=1}[\Phi+i\eta,A=0] + \half \int\frac{d^2x}{\pi}
(\d_\mu\eta)^2.
\label{anu1}
\fin

This equation encodes the anomalous transformation law of the
determinant $Det[i\dsl+\Asl]$ under a chiral gauge transformation
of the abelian gauge field $A$~:
\begin{eqnarray}
\frac{Det[i\dsl+\Asl]}{Det[i\dsl]}=
\exp\[{-\half\int ~\frac{d^2x}{\pi}~(\d_\mu\eta)^2 }\]
\non
\end{eqnarray}

The fact that the field $\eta$ can be
absorbed into a shift of $\Phi$ does not mean that the quenched
correlation functions are identical to those in the pure
system. Using eq.(\ref{anu1}), we have
\debut
\vev{ \prod_n e^{i\al_n\Phi(x_n)} }_{A_\mu=\ep_{\mu\nu}\d_\nu\eta}
= e^{\sum_n \al_n\eta(x_n)}~~
\vev{ \prod_n e^{i\al_n\Phi(x_n)} }_{A=0}.\non
\fin

It factorizes into the product of the correlation functions
in the pure system times a simple function of the impurities.
However, the average of this function is not irrelevant since
the variables $\eta$ have long-range correlations~:
${\bar {\eta(x)\eta(y)}}= -\pi\sig (\d_\mu\d_\mu)^{-1}(x,y)
= -\sig \log|x-y|$. In particular, it changes the values
of the critical exponents. The dimensions $\De$ of the vertex
operators $\exp(i\al\Phi(x))$ in the quenched theory are~:
\debut
\De_{quenched}^\al = \De_{pure}^\al - \sig \al^2. \non
\fin

Averages of product of correlation functions can be computed
similarly. One has~:
\begin{eqnarray}
{\bar {\vev{\prod_n e^{i\al_n\Phi(x_n)} }_A \cdots
\vev{\prod_m e^{i\beta_n\Phi(y_m)} }_A}}  &=&
\vev{\vev{e^{\sum_n\al_n\eta(x_n)+\cdots+
\sum_m\beta_m\eta(y_m)}}}~\times \non\\
& ~& \times~\vev{\prod_n e^{i\al_n\Phi(x_n)} }_{A=0}~\cdots~
\vev{\prod_m e^{i\beta_n\Phi(y_m)} }_{A=0}
\non
\end{eqnarray}
where $\vev{\vev{\cdots}}$ refers to the $\eta$-correlation
functions with the free field measure (\ref{meseta}).
Clearly, conformal invariance is unbroken in
the random abelian case. More details will be described
elsewhere \cite{DB}.

This way of solving this very simple model is closely related to
the change of variables used in ref.\cite{BrOr}.
What we are going to present is the non-abelian
analogue of this construction. In the following, we will
restrict ourself to the $SU(N)$ sector of $U(N)$.

\bigskip
\noindent {\it iii) Symmetries and Ward identities.}

The currents $J^a_\mu =(\bar \Psi \gamma_\mu t^a \Psi)$
are conserved in the pure system, i.e
$\vev{(\d_\mu J^a_\mu) ~ \CO}_{A=0}=0$. It is well known
in conformal field theory that in the pure system these
currents represent a Kac-Moody algebra, alias a current algebra, of
level one \cite{BaHa}.
This is non longer true in the presence of impurities.
However, the measure $P[A]$, eq.(\ref{pa}), is invariant
under a global $SU(N)$ rotation~: $A_\mu\to UA_\mu U^{-1}$
with $U\in SU(N)$. The Noether current associated to this
symmetry is $A_\mu$, which is therefore conserved inside
any quenched correlation functions. I.e.
${\bar {(\d_\mu A_\mu )~ \vev{\CO_1}_A\cdots \vev{\CO_M}_A}}=0$.
By integration by part, this is equivalent to the conservation
law for the currents $\CJ^a_\mu$ defined by~:
\debut
\CJ^a_\mu = \pi \frac{\de}{\de A^a_\mu}. \label{Ccur}
\fin
Explicitely, the Ward identities are~:
\debut
{\bar {\d_\mu~\({ \CJ^a_\mu
\[{ \vev{\CO_1}_A\cdots \vev{\CO_M}_A}\] }\)}} = 0.\label{ward}
\fin
Equivalently, by computing the action of $\CJ^a_\mu$ on
the expectation values, we get~:
\debut
\sum_j {\bar { \vev{\CO_1}_A\cdots
\[{\vev{\d_\mu J^a_\mu \CO_j}_A - \vev{\d_\mu J^a_\mu}_A\vev{\CO_j}_A}\]
\cdots \vev{\CO_M}_A }} = 0.\non
\fin
%Computing the action of $\CJ^a_\mu$ on
%an expectation value, we get~:
%\debut
%\CJ^a_\mu~\vev{\CO}_A = \vev{J^a_\mu O}_A - \vev{J^a_\mu}_A\vev{\CO}_A\non
%\fin
Therefore, in the quenched theory the conserved currents are not
$J^a_\mu$ but their insertions in the connected Green functions.
This point can also be recovered using the replica formalism.
(There also exist Ward identites for chiral gauge transformations~:
$A_z\to UA_zU^{-1}$ and $A_\zb\to VA_\zb V^{-1}$).
As usual, the Ward identites (\ref{ward}) are up to
contact terms which encode the transformation law of the fields.

Usually, conformal field theories with a global $SU(N)$ symmetry
group possess two chiral sets of $SU(N)$ conserved currents
satisfying a Kac-Moody algebra. So, it seems reasonable to expect
that the field theory describing the large distance behavior of
this quenched model should possess two sets of currents.
We will verify this point in the following section while
computing the chiral quenched correlation function.
We will actually obtain a little more since we will find that
the {\it chiral} correlation functions are purely algebraic
and do not present any crossover between their ultraviolet
and infrared behaviors.

\section{The supersymmetric approach.}

A standard approach for studying this random model would be
to use the replica trick or the supersymmetric formalism.
In this section, we use a perturbative renormalization
group computation to derive the infrared behavior
of a few averaged correlation functions.

\bigskip
\noindent {\it i) The supersymmetric action.}

Since the partition function is a determinant, we choose to
use the supersymmetric formalism. Therefore, we introduce
auxiliary bosonic fields, denoted $(b_j,c^j)$, in order to
represent the inverse of the partition function (\ref{part})~:
\debut
\inv{Z[A]}=\int DbDc ~~\exp(-S_{aux}[A]) \non
\fin
with
\debut
S_{aux}[A] = \int \frac{d^2z}{\pi}
\({b_j\( \d_\zb \de^j_k + A^j_{\zb,k}\)c^k +
\bar b_j\( \d_z \de^j_k + A^j_{z,k}\)\bar c^k }\)
\label{saux}
\fin
In order to calculate averages of products of correlation
functions, we need to introduce as many copies of the
fermions and the auxiliary bosons as necessary.
We denote them by $(\bar \Psi^\al,\Psi^\al)$ and
$(b^\al,c^\al)$. Their dynamic is governed by the
action $S_{tot}[A]$ with~:
\debut
S_{tot}[A] = \sum_\al\({ S^\al[A] + S_{aux}^\al[A]}\) .
\label{stot}
\fin
The fields are only coupled through the impurities. After
integration over $A$ with the measure (\ref{pa}) we obtain
the effective action~:
\debut
S_{eff}= \sum_\al \({ S^\al[A=0] + S_{aux}^\al[A=0]}\)
 - \sig  \sum_a\int \frac{d^2z}{\pi}
(\sum_\al H^{a;\al}_z)~(\sum_\al H^{a;\al}_\zb) .\label{seff}
\fin
where $H^a_z$ and $H^a_\zb$ denote the total $SU(N)$ currents~:
\debut
H^a_z &=&  \psi_{-;j} (t^a)^j_k \psi^k_+ + b_j (t^a)^j_k c^k \non\\
H^a_\zb &=& \bar \psi_{-;j}  (t^a)^j_k \bar \psi^k_+ +
\bar b_j (t^a)^j_k \bar c^k. \non
\fin

The action $S_{eff}$ is explicitely supersymmetric. More precisely,
the action $S_{tot}[A]$ is supersymmetric before any integration
over $A$. There are two supercharges, $Q$ and $\bar Q$, acting
separately on the left and right sectors. The charges $Q$, acting
only on the left movers, is defined by~:
\debut
Q(\psi_-) =   0 \quad &;&\quad Q(b)= \psi_- ,\non\\
Q(\psi_+) =  c \quad &;&\quad Q(c)= 0 .\non
\fin
$Q$ acts trivially on $A$ and on the right sector. Clearly, it
satisfies $Q^2=0$. One verifies that $S_{tot}[A]$ is susy invariant~:
$Q(S_{tot}[A])=\bar Q(S_{tot}[A])=0$.  One also verifies that the
total current $H^a_z$ and $H^a_\zb$ are respectively $Q$ and $\bar Q$
closed. This means that there exist two local fields $K^a_z$ and $K^a_\zb$
such that~:
\debut
H^a_z=Q(K^a_z) \quad and\quad H^a_\zb=\bar Q(K^a_\zb).\non
\fin
This property ensures that the total partition function, obtained
by integrating over all the fields (the fermions plus the bosons),
is independent of $A$ as it should be.

There exists an intriguing similarity between the above construction and
conformal topological field theory in two dimensions. In the latter,
one uses a twisted version of a $N=2$ supersymmetry to construct the
theory such that the stress tensor is susy closed. The closure of
the stress tensor ensures that the partition function of the
topological theory is independant of the background metric.
In the random model, we can extend the supersymmetric algebra we
just described such that the new algebra possesses two supersymmetric
charges. The fact that the current is susy closed is the analogue
of the fact that the stress tensor is susy closed.
It seems natural to wonder if techniques of conformal topological
field theories cannot be translated into new methods of analysis
of two dimensional random systems.

\bigskip
\noindent {\it ii) A renormalisation group computation.}

The only coupling constant which can be renormalized is $\sig$.
As usual, the one-loop $\beta$ and $\ga$-functions are encoded
in the operator product expansions of the fields in the unperturbed
theory. These can be read off from the action $S_{tot}[A=0]$. We
have~:
\debut
\psi^k_+(z)~\psi_{-;j}(w) \sim \frac{\de^k_j}{(z-w)} + regular,\non\\
c^k(z)~b_j(w) \sim \frac{\de^k_j}{(z-w) }+ regular.\non
\fin
The total currents $H^a_z$ satisfy a Kac-Moody algebra with zero
central charge~:
\debut
H^a_z(z)~H^b_z(w) \sim \frac{if^{abc}}{z-w} H^c_z(w) + reg.\non
\fin
There is an exact compensation between the central charges associated
to the fermions and to the auxiliary bosons.

The currents $H^a_z$ have dimension one; therefore, the $\beta$
function vanishes at tree level. At one loop, one finds~:
\debut
\dot \sig \equiv \beta(\sig) = - C_G~ \sig^2 +\CO(\sig^3),
\label{betaf}
\fin
where $C_G>0$ is the Casimir in the adjoint representation,
$f^{abc}f^{dbc}=C_G\de^{ad}$. The sign in eq.(\ref{betaf}) is
important; it tells us that the model is asymptotically free
in the infrared regime. We have~:
\debut
\sig(R) = \frac{\sig_0}{1 + \sig_0 C_G \log(R/a)} ,\non
\fin
with $\sig_0>0$ the value of the coupling constant at the lattice cut-off $a$.
Since the model is asymptotically free in the infrared, we can use the
renormalization group to evaluate the large distance behavior of
the two-point correlation functions using the usual formula~:
\debut
\vev{\Phi(R)\Phi(0)}_{\sig_0} = \vev{\Phi(a)\Phi(0)}_{\sig(R)}
\exp\({-2\int_{\sig_0}^{\sig(R)} d\bar \sig
\frac{\ga_\Phi(\bar \sig)}{\beta(\bar \sig)} }\),
\fin
where $\ga_\Phi$ is the $\ga$-function for $\Phi$.

Let us first compute the asymptotic of the chiral two-point
function ${\bar {\vev{J^a_z(z)J^b_z(0)}_A}}$.
To evaluate it we only need one copy of the fermion
and their supersymmetric partner. We find that the $\ga$-function
of $J^a_z$ is $\ga_J(\sig)= 1 + \CO(\sig^2)$; i.e.
there is no one-loop correction even though $J^a_\mu$ are not conserved.
Therefore, at large distances we have~:
\debut
{\bar {\vev{J^a_z(z)J^b_z(0)}_A}}
\sim \frac{\de^{ab}}{z^2} .\label{rg}
\fin
In particular, there are no logarithmic correction.

Consider now the average of the product of two such two-point
functions but of opposite chiralities, i.e.
${\bar {\vev{J^a_z(z)J^b_z(0)}_A
\vev{J^{\bar a}_\zb(z)J^{\bar b}_\zb(0)}_A}}$.
To compute it we need two copies of the fermions. These quenched
averages are then represented by the two-point functions of the
operators $\CO_{12}^{a \bar a}=
J^{a;(1)}_z(z)J^{\bar a;(2)}_\zb(z)$,
products of currents in the different copies.
For these operators the $\ga$-functions have one-loop corrections.
There is a mixing between the operators $ \CO_{12}^{a \bar a}$
with different indices~:
\debut
\ga^{a \bar a}_{b \bar b}= 2\de^a_b\de^{\bar a}_{\bar b}
+2\sig  f^{d a}_{b}f^{d\bar a}_{\bar b} +\CO(\sig^2).\non
\fin
In particular the $SU(N)$ scalar operator $\CO=\sum_a
J^{a;(1)}_zJ^{a;(2)}_\zb$ does not mix with the others, and
$\ga_\CO= 2 + 2C_G + \CO(\sig^2)$. Therefore,
\debut
\sum_{a,b}
{\bar {\vev{J^a_z(z)J^b_z(0)}_A\vev{J^a_\zb(z)J^b_\zb(0)}_A}}
\sim \({\frac{a^2}{z\zb}}\)^2
\({\log|z\zb|}\)^{-4} \label{logcor}
\fin
Notice the logarithmic correction which breaks the self-averaging
property. It is remarkable that the logarithms appear in the averages
of products of correlations of opposite chiralities, eq.(\ref{logcor}),
but not in the averages of the chiral correlations, eq.(\ref{rg}).

Infrared asymptotics of averages of higher products of current
two-point functions, or of fermion correlations, can be
computed similarly. In particular, we find logarithmic
corrections for the quenched two-point function
${\bar {\vev{\ep(R)\ep(0)}_A}}$ for the energy operator
$\ep= \bar \Psi \Psi$~:
\debut
{\bar {\vev{\ep(z)\ep(0)}_A}}\sim \({\frac{a^2}{z\zb}}\)~
\({\log|z\zb|}\)^{-4C_V/C_G}. \non
\fin
where $C_V$ is the $SU(N)$ Casimir in its $N$-dimensional
representation.

\section{The quenched chiral correlation functions.}

In this section, we describe how some of the quenched correlation
functions can be directly computed without using the replica
trick or the supersymmetric formulation.
This relies on the Polyakov-Wiegman (PW) formula \cite{PW}
for the effective action $W[A]$, eq.(\ref{effec}).
Once this formula has been recalled, the computations
are rather simple.

\bigskip
\noindent {\it i) The Polyakov-Wiegman formula.}

The PW formula relies on the fact that in two dimensions the
effective action can be exactly computed by integrating
its anomalous transformation under a chiral gauge transformation.
This is the non-abelian analogue of eq.(\ref{anu1}).
Indeed, let $G_\mu=\sum_at^aG_\mu^a[A]$
be the generating functions of the
connected current Green function in the pure system~:
\debut
G_\mu^a[A] \equiv \vev{J_\mu^a}_A = \pi\frac{\de W[A]}{\de A^a_\mu}.
\label{onep}
\fin
They satisfy the anomalous Ward identities \cite{PW},
\debut
\d_z G_\zb + \d_\zb G_z +[A_z,G_\zb]+[A_\zb,G_z] &=& 0, \non\\
\d_z G_\zb - \d_\zb G_z +[A_z,G_\zb]-[A_\zb,G_z] &=&
2 F_{z\zb}[A] , \label{ano}
\fin
with $F_{z\zb}[A]=\d_z A_\zb - \d_\zb A_z +[A_z,A_\zb]$.
Eqs.(\ref{ano}) completely specify $G_\mu[A]$~:
\debut
 G_z[A] &=& A_z - \inv{\d_\zb + ad.A_\zb} \d_zA_\zb \label{ga}\\
 G_\zb [A] &=& A_\zb - \inv{\d_z + ad.A_z} \d_\zb A_z \label{gza}
\fin
Here $ad.A_z$ denoted the adjoint action of $A_z$. Notice that
$G_z[A]$ is  local in $A_z$ but non-local in $A_\zb$.

\bigskip
\noindent {\it ii) Explicit formula for the chiral correlation functions.}

Knowing explicitely the correlation functions for any impurity sample,
it is a priori possible to take the quenched average. But let
us first concentrate on the average of the chiral correlation
functions involving only currents of the same chirality; e.g.
involving only $J^a_z$. Consider first the average of products of
one-point functions (\ref{onep}). Since the quenched average
is defined by ${\bar {A_z^a(x)A_\zb^b(y)}}=\sig\pi\de^{ab}\de^{(2)}(x-y)$
and since $\vev{J_z^a}_A$ are linear in $A_z$, these quenched
correlations can be computed using the Wick theorem applied on $A$.
For exemple, the two-point and three-point functions are~:
\debut
{\bar {\vev{J_z^a(z_1)}_A\vev{J_z^b(z_2)}_A}}
&=& 2\sig\pi~\de^{ab}~ \({\inv{\d_\zb}~\d_z}\)_{(z_1,z_2)}
= \frac{ 2\sig \de^{ab} }{ (z_1-z_2)^2} \non\\
{\bar {\vev{J_z^a(z_1)}_A\vev{J_z^b(z_2)}_A\vev{J_z^c(z_3)}_A}}
&=& i(\pi\sig)^2 f^{abc} \({\inv{\d_\zb}~\d_z}\)_{(z_1,z_2)}
\[{\({\inv{\d_\zb}}\)_{(z_3,z_1)}-\({\inv{\d_\zb}}\)_{(z_3,z_2)}}\]\non\\
 && ~~~~ + (cyclic~~ permutation) \non\\
&=& 3\sig^2 \frac{if^{abc}}{(z_1-z_2)(z_1-z_3)(z_2-z_3)} \non
\fin

More generally, the average of products of one-point functions
is the sum of connected correlations which can be expressed
in terms of the correlation functions of the pure system~:
\debut
\[{ {\bar {\vev{J_z^{a_1}(z_1)}_A \cdots \vev{J_z^{a_M}(z_M)}_A }}
}\]^{connected} = M \sig^{M-1} \vev{J_z^{a_1}(z_1) \cdots J_z^{a_M}(z_M)}_0
\label{avone}
\fin
Here, $\vev{\cdots}_0$ denote the pure correlation functions.
They are known exactly \cite{KZ}. The relation between the quenched
correlations and their connected parts is the usual one.

More interesting is the average of products of correlations
with insertion of the conserved currents (\ref{Ccur}) since they
encode the underlying symmetry algebra. We find~:
\debut
 &&{\bar {\CJ_z^{n_1}(z_1) \cdots \CJ_z^{n_M}(z_M)
\vev{J_z^{a_1}(w_1)J_z^{b_1}(\xi_1)\cdots}^c_A\cdots
\vev{J_z^{a_P}(w_P)J_z^{b_P}(\xi_P)\cdots}^c_A }} \non\\
&=& \vev{J_z^{n_1}(z_1)\cdots J_z^{n_M}(z_M)
J_z^{a_1}(w_1)J_z^{b_1}(\xi_1)\cdots}_0\cdots
\vev{J_z^{a_P}(w_P)J_z^{b_P}(\xi_P)\cdots}_0  \non\\
&+& \sum_{j=1}^M
\vev{J_z^{n_1}(z_1)\cdots {\widehat {J_z^{n_j}(z_j)}}
\cdots  J_z^{n_M}(z_M)
J_z^{a_1}(w_1)J_z^{b_1}(\xi_1)\cdots}_0\times\label{bigcor}\\
&& ~~~~~~\times
\vev{J_z^{n_j}(z_j) J_z^{a_2}(w_2)J_z^{b_2}(\xi_2)\cdots}_0
\cdots \vev{J_z^{a_P}(w_P)J_z^{b_P}(\xi_P)\cdots}_0  \non\\
&+&\cdots  \non\\
&+& \vev{J_z^{a_1}(w_1)J_z^{b_1}(\xi_1)\cdots}_0\cdots
\vev{J_z^{n_1}(z_1)\cdots J_z^{n_M}(z_M)
J_z^{a_P}(w_P)J_z^{b_P}(\xi_P)\cdots}_0 \non
\fin
The hatted fields have to be omitted.
Here, we assumed that there is no insertion of one-point functions.
The formula (\ref{bigcor}) is actually simpler in words~:
it is obtained by distributing the currents $J_z^{n_1}(z_1),
\cdots,J_z^{n_M}(z_M)$ among the pure correlators in all
possible way, each counted only once.

Notice that all the chiral quenched correlation functions are
purely algebraic, without any logarithmic correction, in
agreement with the renormalization group computation.
So, conformal invariance does not seem to be broken in
the random model.

 From eq.(\ref{bigcor}) we read the operator product expansion
of the fields. The currents satisfy~:
\debut
\CJ_z^{n_1}(z_1) \CJ_z^{n_2}(z_2)
= \frac{\de^{n_1n_2}}{(z_1-z_2)^2}
+ \frac{if^{n_1n_2n_3}}{z_1-z_2} \CJ^{n_3}(z_2) + reg. \label{ope1}
\fin
Therefore, the quenched conserved currents satisfy the
commutation relations of a Kac-Moody algebra, exactly
as the currents in the pure system do.

The operator product expansion between the conserved currents
and the correlators $\vev{J_z^{a_1}(w_1)J_z^{a_2}(w_2)\cdots}_A$ are~:
\debut
\CJ_z^{n}(z) \vev{J_z^{a_1}(w_1)J_z^{a_2}(w_2)\cdots}^c_A
&=& \sum_j \frac{\de^{na_j}}{(z-w_j)^2}
\vev{J_z^{a_1}(w_1)\cdots {\widehat {J_z^{a_j}(w_j)}}\cdots}^c_A \non\\
&+& \sum_j \frac{if^{na_ja'_j}}{z-w_j}
 \vev{J_z^{a_1}(w_1)\cdots J_z^{a'_j}(w_j)\cdots}^c_A
 + reg.\label{ope2}
\fin
These operator product expansions are unusual in conformal field
theory with Kac-Moody symmetry. In particular, they imply
that the fields  $\vev{J_z^{a_1}(w_1)J_z^{a_2}(w_2)\cdots}^c_A$
are {\it not} associated to highest weight vector representations.
We don't know to which category of representations they
correspond to.

%The quenched correlation functions can be reconstructed from these
%operator product expansions.

Contrary to the chiral quenched correlation functions which are easy to
compute, the averages of correlation functions involving fields
of opposite chiralities are difficult to evaluate.
This is due to the fact that the generating functions $G_z[A]$
and $G_\zb[A]$ are non-local in $A_\zb$ and $A_z$ respectively.
A naive perturbative expansion is spoilled by untractable divergences.
This could have been anticipated in view of the logarithmic
corrections present in eq.(\ref{logcor}). The occurence of
these logarithms probably means that the
gluing of the left and right sectors is more subtle than usual.
In particular, one has to learn how to deal with an infinite number of
primary fields.
We don't know any good algebraic way of doing it correctly
in the non-abelian case, but
we hope that the remarks presented in this note will be useful
to find the answer.

\bigskip
\bigskip

{\bf Acknowledgement.} It is a pleasure to thank K. Gawedzki and
J.-B. Zuber for interesting and stimulating discussions, and
H. Orland for suggesting me to write this note.

While finishing writing this note, a related paper appears
on the bulletin board \cite{Wen}.

\vfill \eject

\section{Appendix.}

In this appendix, we explain why our results seem to be in
contradiction with some of those presented in ref.\cite{Tet}.
The approach used in this paper
is based on the replica trick. Thus, the authors introduce $n$ copies
of the $N$ fermions, forming alltogether $Nn$ Dirac fermions.
These fermions form a representation of level one of
the $SU(Nn)$ Kac-Moody algebra.
After averaging the random potential, these $Nn$ free massless
Dirac fermions are coupled by a current-current interaction~:
\debut
S_{int} = \sig \int \frac{d^2z}{\pi} (\sum_p J^{a;p}_\mu)
 (\sum_p J^{a;p}_\mu) \label{actrep}
\fin
where the currents $J^{a;p}_\mu$ are the $SU(N)$ currents in the
$p^{th}$ replica. Not all the degrees of freedom are coupled by
this interaction since only the diagonal sum of the $SU(N)$ currents
interacts. The authors of ref.\cite{Tet} argue that the interaction
opens a gap and that some of the modes become massive.
More precisely, let us decompose the $SU(Nn)_{k=1}$ representation
as $SU(Nn)_{k=1}=SU(N)_{k=n}\times SU(n)_{k=N}$.  They argue that
the modes corresponding to $SU(N)_{k=n}$ become massive and therefore
that the infrared behavior is described by  the $SU(n)_{k=N}$
conformal field theory. In particular,
this would implies that the correlation functions of the diagonal
$SU(N)$ currents $(\sum_p J^{a;p}_\mu)$ decrease exponentially at
infinity. But, since these correlation functions at $n=0$ give the
quenched averages of the connected correlation functions of the currents,
the latter statement is in contradiction with the result
we obtained using the Polyakov-Wiegmann formula.

\vfill \eject

\end{document}